\documentclass[preprint,preprintnumbers,amsmath,amssymb]{revtex4}

\usepackage{graphicx}
\usepackage{amsmath}
\usepackage{dcolumn}

\begin{document}

\title{Reverse magnetic vortex curling direction of ferromagnetic nanodisk}

\author{Wen-Ming Ju, Mark Tuominen\footnote{Author to whom correspondence should be
addressed; electronic mail: tuominen@physics.umass.edu.}}
\affiliation{Physics Department, University of Massachusetts at
Amherst, MA. 01003, USA}
\author{Nihar Pradhan}
\affiliation{National High Magnetic Field Lab, FL. 32310, USA}
\author{Jessica Bickel, Kathy Aidala}
\affiliation{Physics Department, Mount Holyoke College, MA. 01075,
USA}

\date{\today}
 \vskip 2pc
  \vskip 2pc
   \vskip 2pc
    \vskip 2pc

\begin{abstract}

We reverse the magnetic vortex curling direction of ferromagnetic
nanodisk by applying a circular Oersted field. The nanodisk is
fabricated without breaking its symmetry. The Oersted field is
induced by passing current through an atomic force microscope tip
placed at the center of the disk. Micromagnetic simulation indicates
that compared to the uniformly distributed current throughout the
cross section of disk, the line current concentrated in the center
can reverse the chirality more easily, which is in accordance with
our experimental results.

\end{abstract}

\maketitle


Ferromagnetic nanodisks have attracted attention over the past few
years due to its unique closed-flux vortex
state~\cite{Shinjo,Scholz,Kasai,Yamada,Guslienko}. The curling
direction of the in-plane vortex state is either clockwise or
counter-clockwise. This degree of freedom is defined as
\emph{chirality}. In the center, the magnetization pops out,
pointing either up or down, and we denote this degree of freedom as
\emph{polarity}. The resulting four degenerate states are
independent, and have remarkably stability against thermal
fluctuations. The control of both polarity and chirality could
stimulate the use of vortex state in nonvolatile data storage and
random access memory devices. Polarity switching has been realized
in a lot of studies by using various techniques, including
out-of-plane perpendicular current or magnetic
filed~\cite{Caputo,Sheka}, in-plane spin-polarized
current~\cite{Yamada2,Yamada} and magnetic
field~\cite{Waeyenberge,Hertel,Yu}. However, it is still hard to
manipulate the second degree of freedom due to the high symmetry of
the structure. In some studies, the nanodisk is modified somehow to
break the symmetry, so the chirality is able to be switched in the
nanodisk with notch, truncated edge or with asymmetrical magnetic
properties in lateral
direction~\cite{Kimura,Schneider,Zhong,Cambel}.

We present here that we can switch the chirality without breaking
the symmetry of the nanodisk. We realize this by applying a circular
Oersted field that is induced by passing current through an atomic
force microscope tip placed at the center of the disk, and we get
the information of curling direction by magnetic force microscopy
(MFM) images of the initial and final magnetization states with an
in-plane uniform magnetic field applied. Micromagnetic simulation is
also conducted to help explain the underlying mechanism.

Permalloy nanodisks are fabricated on a gold-coated silicon wafter
by using standard electron-beam lithography and lift-off process.
The diameter and thickness of the disk are 1 $\mu$m and 50 nm,
respectively, to ensure the vortex magnetization distribution for
the ground state. The magnetic dipole moments have the in-plane
closed form around the center in the vortex state, so MFM image
presents no contrast for the disk except the center, and one can not
tell the curling direction for the time being. However, if we apply
an in-plane uniform magnetic field, it is expected that more and
more magnetic moments will be aligned with the magnetic field to
minimize Zeeman energy, then the vortex core will be "pushed" away
from the center to the edge of the disk. Furthermore, if the disk
has an opposite chirality, the core will move to the edge on the
other side. So by knowing the direction of the external uniform
magnetic field and observing the moving direction of the vortex
core, we learn the chirality information. Fig. 1 shows the
simulation and experimental verification of this conception. We can
clearly see, from the MFM images (1st and 2nd row) that with the
in-plane uniform magnetic field increased from 0 to 400 Gauss, the
vortex cores of the two disks move from center to the right and left
edges step by step, respectively, showing that they have clockwise
and counter-clockwise curling directions, respectively. The third
row is the simulation results which simulate the process of the
second row. Experimental and simulation results are in good
accordance with each other.

Since the curling direction is known, we can then try to reverse the
chirality. The idea is to apply current to the center of disk
vertically via atomic force microscope (AFM) tip. If the
current-induced circular magnetic field has the opposite curling
direction to that of the disk and is large enough, then the curling
direction of the magnetic moments in the disk is expected to be
reversed. The experimental setup is shown in Fig. 2. The platinum
AFM tip is in contact with the disk in the sample. The sample is
connected to a resistor with resistance $R_0=40$ $\Omega$ by copper
wire and is grounded. We apply voltage $V$ to the circuit, ramp it
up from 0 to 8 Volts at a constant rate, and read out the response
voltage of the resistor $V_0$. We can simply calculate the current
according to $I=V_0/R_0$ and get a plot of $I$ versus $V$. When
applying the voltage, no external filed is applied in order for the
magnetization distribution of the disk to be in vortex-state. Before
and after ramping up the voltage, 600 Gauss in-plane uniform
magnetic field is applied to get chirality information by MFM scan.

With the sample structure illustrated in Fig. 3(a), there is always
a linear relationship between $I$ and $V$ (Fig. 3(b)), and the
chirality can not be reversed, as MFM images captured before and
after the voltage application show no difference. However, after the
sample structure is slightly altered, the result turns out to be
what is expected. This time, based on the previous sample, a 10 nm
thick Ti layer is deposited on top of Permalloy disk. Then a thin
layer of TiO$_2$ is grown on Ti layer via plasma-enhanced chemical
vapor deposition (PECVD) (Fig. 3(c)). The following MFM measurement
and voltage application process are the same as before. The $I-V$
curve is shown in Fig. 3(d) (note that this time the voltage is
ramped up from 0 to -8 Volts because the disk has an opposite
chirality as compared to the previous one). As we can see,
initially, the current is almost 0 as the voltage is ramped up. At
some voltage $V_B$, the current suddenly jumps from 0 to $V_B/R_0$.
Afterwards, it increases linearly as the voltage increases, with a
slope the same as that in Fig. 3(b). This phenomenon is explainable:
When the "jump" happens, the TiO$_2$ layer as a dielectric is
penetrated so that the disk transforms from an insulator to a
conductor. Experimental trials show that everytime the voltage is
applied to the disk with the Ti and TiO$_2$ capping layers, a jump
appears in the $I-V$ curve. In addition, when the breakdown voltage
$V_B$ is above some value (approximately 5 Volts), the chirality can
be reversed since the vortex core is at opposite sides of the disk
before and after the voltage application with the same in-plane
uniform magnetic field applied (Fig. 4, disk 1 and 2). When $V_B$ is
below 5 Volts, the vortex core presents at the same side (Fig. 4,
disk 3), illustrating that the chirality is not reversed.

To explain the phenomenon that only a jump appearing at voltage
above 5 Volts approximately in the $I-V$ curve can lead to chirality
switch, we conduct micromagnetic simulation based on OOMMF. In the
simulation, we assume two cases of current distribution: 1.
Uniformly distributed all over the cross section of the disk, then
according to Ampere's law, magnetic field $B\propto Jr$, where $J$
is current density, $r$ is the distance between disk center and
point investigated; 2. Concentrated in the center, like a
line-current, so $B\propto I/r$. We then calculate the magnetic
field at each simulation cell and generate field file to run
simulation for the two current-distribution cases with various $I$
and $J$. The results show that in the uniform distribution case, the
threshold current density $J_t$ for the chirality to switch is
$1\times10^{12}$ A/m$^2$. The corresponding threshold current is
$I_{t1}=\pi R^2J_t=0.64$ A where $R=450$ nm is the radius of disk,
whereas in the line distribution case, in order for the chirality to
switch, the threshold current is $I_{t2}=0.03$ A, which is much
smaller than $I_{t1}$. Consequently, we think that for disk without
dielectric capping layer, when $I$ is steadily increasing with $V$,
the current spreads out when going through the disk, and the extreme
situation for this scenario is the uniform distribution case.
However, the maximum current applied in our experiment is 0.20 A,
which is smaller than the threshold current $I_{t1}$ for this case,
so the chirality can not be reversed. On the other hand, for disk
with dielectric capping layer, the current increases a lot within a
very short time when the dielectric is penetrated, thus the current
is more concentrated in the disk, and more like a line current. In
other words, when we compare the two points in Fig.3 (denoted as
Point 1 and 2) with the same magnitude of current and resort to the
equation $I=envA$, where $I$ is the current flowing in disk, $e$ the
charge of an electron, $n$ the charge density, $v$ the drift speed,
$A$ the cross-sectional area, we know that $I$, $e$, and $n$ are the
same for the two points. However, $A$ is much smaller in the second
point as $v$ is much larger in the breakdown case. Accordingly,
current is more concentrated. Simply speaking, the dielectric
penetration creates a conducting channel that is much narrower than
that formed by direct contact between MFM tip and ferromagnetic
disk. When the jump happens at $V_B\simeq5$ Volts, the current
immediately after the jump is $I=V_B/R_0=0.125$ A. This is larger
than the threshold value $I_{t2}$ for the line distribution case, so
the chirality is able to be reversed. Moreover, when the jump
happens at $V_B<5$ Volts, the chirality cannot be reversed, meaning
that the actual threshold current is about 0.125 A, larger than
$I_{t2}$. This is reasonable because although the current is
distributed more concentratively after the breakdown of the
dielectric layer, it is not an ideal line current. Thus, larger
threshold current is required to switch the chirality.

In summary, we have demonstrated magnetic vortex chirality switch of
ferromagnetic nanodisk. This is realized by applying a current
induced circular Oersted field. Simulation shows that the current
density needs to be at least $1\times 10^{12}$ A/m$^2$ to accomplish
the switch in the case that the current is uniformly distributed
throughout the cross section of the disk. The corresponding current
is about 0.6 A. The actual current applied to switch the chirality
in our experiment is no more than 0.2 A. According to the
simulation, we think that our capacitor-like sample is able to
concentrate the current enormously and thus less current is required
to realize the switch. We hope our study can bring fresh notion for
magnetic states manipulation and relevant application.

This work was supported by NSF grants DMR-1208042, DMR-1207924 and
CMMI-1025020.

 \vskip 2pc
  \vskip 2pc
   \vskip 2pc


\clearpage

\begin{figure}[h]
\includegraphics[width=0.85\textwidth]{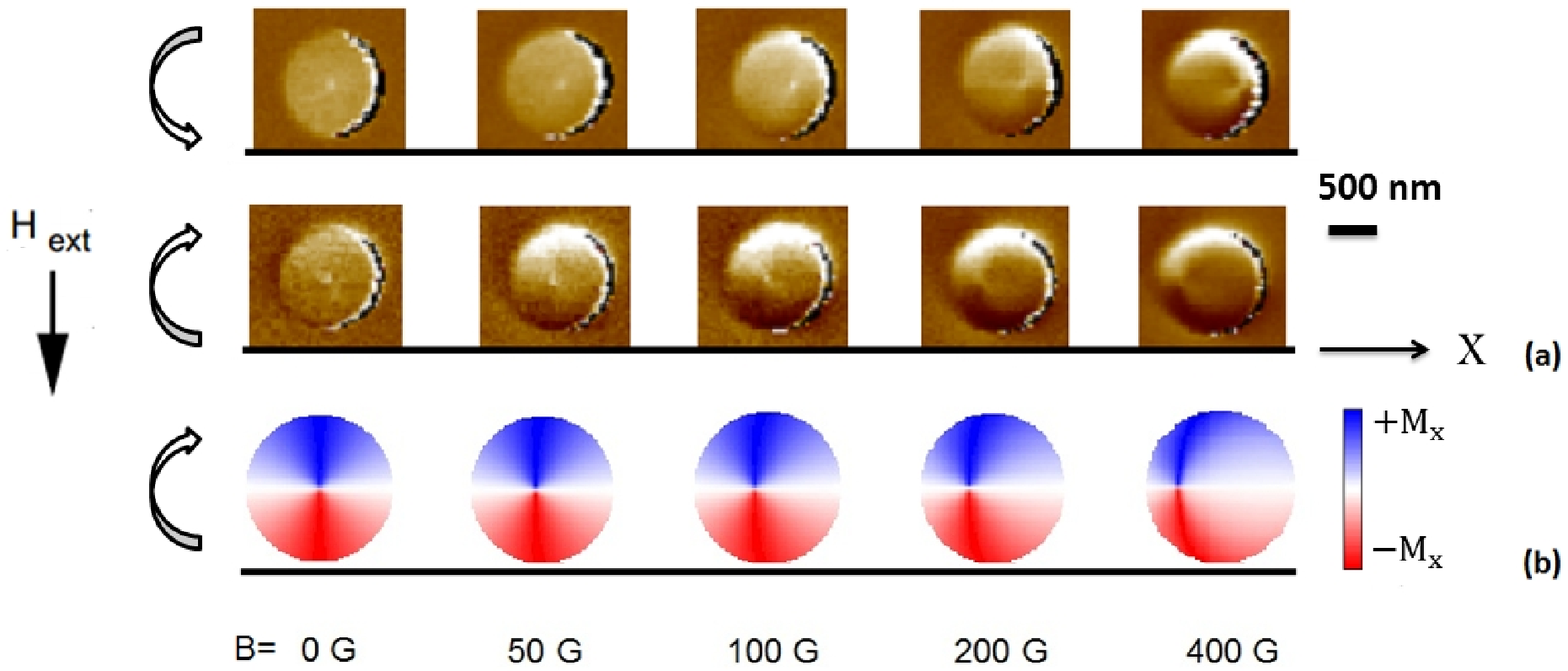}
\caption{Experimental and simulation verification of vortex core
motion. Each column corresponds to one specific external uniform
magnetic field applied in the sample plane with its magnitude shown
on the bottom. (a). MFM images showing vortex core motion of a
nanodisk with counter-clockwise (1st row) and clockwise (2nd row)
curling directions, respectively, as shown by the curling arrows.
(b). Magnetization images from OOMMF simulation showing the process
of the second row in (a).\label{your label}}
\end{figure}

\begin{figure}[h]
\includegraphics[width=0.5\textwidth]{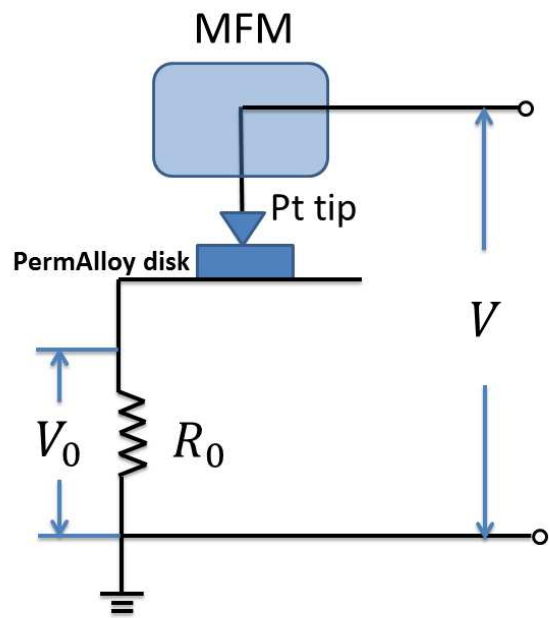}
\caption{Schematic illustration of the experimental
setup.\label{your label}}
\end{figure}

\begin{figure}[h]
\includegraphics[width=1.0\textwidth]{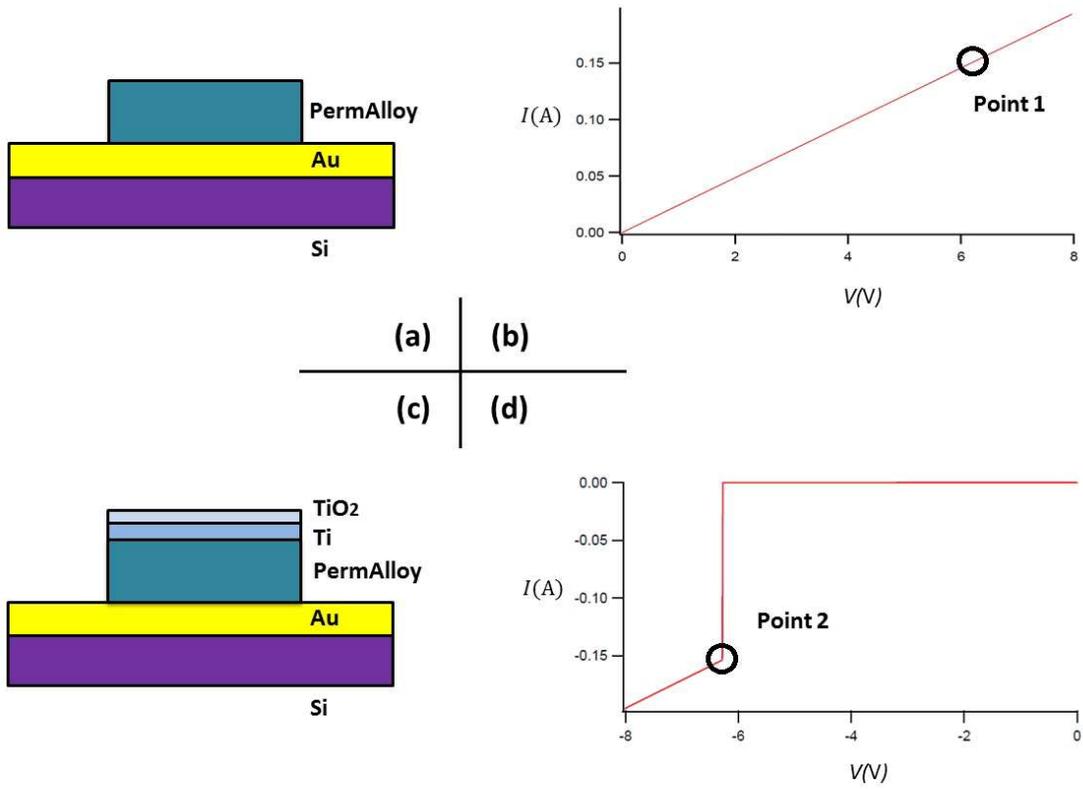}
\caption{(a), (c). Schematic illustrations of the sideview of two
different disk structures investigated in our study. (b), (d). $I-V$
curves for samples (a) and (c), respectively.\label{your label}}
\end{figure}

\begin{figure}[h]
\includegraphics[width=0.9\textwidth]{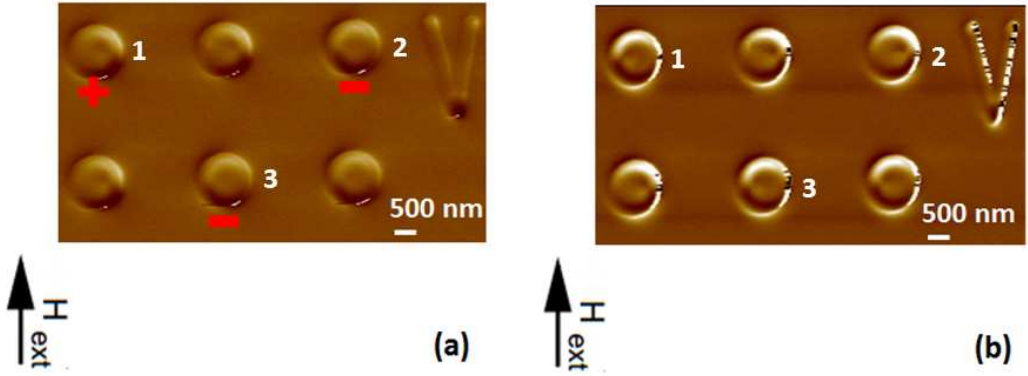}
\caption{MFM images of six nanodisks before (a) and after (b)
applying current to three of them as indicated by "1", "2" and "3".
The same external uniform magnetic field is applied to both images
in sample plane. "+" and "-" signs illustrate positive and negative
voltage application, respectively. All the disks have the same
structure as that in Fig. 3(c).\label{your label}}
\end{figure}

\end{document}